\documentclass[12pt]{iopart}
\usepackage{graphicx}
\usepackage{bm}
\usepackage{amssymb}
\usepackage{color}
\usepackage{cite}

\begin{document}

\title{Periodic patterns displace active phase separation
}
\author{Frederik J. Thomsen, Lisa Rapp, Fabian Bergmann, Walter Zimmermann}
\address{Theoretische Physik I, Universit\"at Bayreuth, 95440 Bayreuth, Germany}

\begin{abstract}
In this work we identify and investigate a novel bifurcation in conserved systems. 
This  secondary bifurcation    stops  active phase separation in its nonlinear regime. 
It is  then either replaced by an extended, system-filling, spatially periodic pattern or, in a complementary parameter region, by a novel hybrid state with spatially alternating  homogeneous and periodic states. 
The transition from  phase separation to extended spatially periodic patterns is hysteretic. 
We show that the resulting patterns are multistable, as they show stability beyond the bifurcation
 for different wavenumbers belonging to a wavenumber band.
  The transition from active phase separation to the hybrid states is continuous.
 Both  transition scenarios are systems-spanning phenomena in particle conserving systems. They are
 predicted with a   generic dissipative  model introduced in  this work. 
Candidates for specific systems, in which these generic secondary transitions are likely to occur, are, for example, 
generalized models for motility-induced phase separation in active Brownian particles, models for cell division or chemotactic systems with conserved particle dynamics.    

\end{abstract}

\section{Introduction}

Patterns are ubiquitous in nature. They emerge spontaneously in a plethora of living or inanimate driven systems
    \cite{Ball:98,CrossHo,Cross:2009,Pismen:2006,Aranson:02.1,Lappa:2010,Kondo_Miura:2010.1,Meron:2015,Meron:2018.1,BaerM:2020.1},
and their esthetic appeal   is immediately apparent to all observers \cite{Ball:98}.
 Prototypical patterns, such as stripes, hexagons and traveling waves can be found in various fluid-dynamical, chemical and biological systems and they often play a key role in various processes in nature.    To name just a few examples:
Convection
patterns in fluid dynamical systems enhance the thermal transport \cite{CrossHo,Lappa:2010}, protein dynamics in \textit{E. coli}  plays
 a key role in finding the cell center during the cell division process 
\cite{deBoer:1999.1,Lutkenhaus:2007.1,Schwille:2008.1,Loose:2011.1}, 
  and the formation of patterns are the basis of successful survival strategies for vegetation in water-limited regions \cite{Meron:2015,Meron:2018.1}.

 The mechanisms that drive  nonequilibrium patterns are as diverse as the systems in which they occur. 
Instabilities leading to patterns are usually divided into  type I, II and III,  which differ in their preferred wavenumber  $q_0$ 
and/or in their frequency  $\omega_0$ and the wavenumber dependence  of the dispersion relation  \cite{CrossHo}.
Despite their very different origins,  nonequilibrium patterns of the  type I and type III share  common universal properties, which
 are covered by universal Ginzburg-Landau equations (GLEs)
  \cite{Ginzburg:50.1,Ginzburg:2004.1,Newell:1969.1,Segel:69.1,Newell:1971.1,Stewartson:71.1,Kuramoto:84,CrossHo,Newell:1993.1,Aranson:02.1,Pismen:2006,Cross:2009}.  They describe the dynamics of the envelope (resp. the order parameter field) of
stationary and traveling stripe  or oscillating patterns.

A number of  nonequilibrium demixing phenomena with  conserved particle dynamics attract increasing attention recently
 \cite{EdelsteinKeshet:2013.1,Grill:2014.1,HillenT:2009.1,Romanczuk:2014.1,Palacci:2013.1,Cates:2015.1,SpeckT:2014.1,Marchetti:2016.1,SpeckT:2020.1,Fromherz:1991.1}.  Ultimately, these are
  type II instabilities, for which the theory is currently undergoing a lively development.
Such nonequilibrium demixing phenomena include cell polarization  \cite{EdelsteinKeshet:2013.1,Grill:2014.1},
chemotactically communicating cells \cite{HillenT:2009.1,Romanczuk:2014.1}, motility induced phase separation (MIPS) in self-propelled Brownian particle systems  \cite{Palacci:2013.1,Cates:2015.1,SpeckT:2014.1,Marchetti:2016.1,SpeckT:2020.1} and also a very early model for the ion-channel density in a membrane \cite{Fromherz:1991.1}.

 Demixing phenomena in thermal equilibrium systems are long known and  they are described by the Cahn-Hilliard (CH) mean-field model   \cite{Cahn:58.1,Cahn:1961.1,Bray:1994.1,Desai:2009}.
It is very surprising that the CH equation describes also the generic order-parameter field  for type II instabilities with particle conservation, i.e. also for non-equilibrium demixing phenomena \cite{Bergmann:2018.2,Rapp:2019.1,Bergmann:2019.1,Rapp:2019.2}.
This was shown by generalizing the perturbation theory used to derive equations for unconserved  order parameter fields \cite{Newell:1969.1,Segel:69.1,Newell:1971.1,Stewartson:71.1,CrossHo,Newell:1993.1,Cross:2009} to cases with conserved order parameter fields \cite{Bergmann:2018.2}.
  Its  application to the
dynamical equations of chemotactic systems \cite{Bergmann:2018.2,Rapp:2019.2},
to models of cell polarization \cite{Bergmann:2018.2,Bergmann:2019.1} and to MIPS \cite{Rapp:2019.1}
 results always directly in the generic CH equation.   Accordingly,   the coefficients of the CH equation now  depend on the kinetic parameters  of the respective   system  and it describes 
nonequilibrium rather than equilibrium transitions.
 
In the case of MIPS the CH model was extended  to the active model B and the active model B+ by higher order nonlinear terms to describe interesting coarse-grained phenomena further away from the  onset \cite{Cates:2014.1,Cates:2017.1,Cates:2018.1}  with particular emphasis on the effects of noise.
The perturbation theory in Ref. \cite{Bergmann:2018.2} extended to the next higher nonlinear order and  applied to a mean-field theory for MIPS \cite{SpeckT:2014.1,SpeckT:2015.1} gives the generic structure of AMB+ \cite{Rapp:2019.1} as well. Moreover, this perturbation-theoretic approach unambiguously links the parameters of  AMB+  to the kinetic parameters of the mean field model in Refs.~\cite{SpeckT:2014.1,SpeckT:2015.1}. 
In terms of this perturbation theory  the deterministic part of AMB+ is a generic nonlinear extension of the CH model rather than an extension into nonequilibrium.

  The perturbation theory  from Ref.   \cite{Bergmann:2018.2} systematically provides, in addition to the generic next higher nonlinear contributions, also a sixth-order spatial derivative of the order parameter field.   This higher order derivative is not part of active model B+, but becomes indispensable  beyond a finite distance from the threshold of MIPS and at larger amplitudes of the order parameter field. In this range, solutions without the sixth-order derivative then show  unbounded growth. The  conserved Swift-Hohenberg  model+ (CoSH+)  introduced in section \ref{genmodel} for a class of systems showing nonequilibrium phase separation includes
this higher derivative and shows  a novel secondary bifurcation in the nonlinear regime of 
active phase separation, which is characterized in section \ref{phasepatt}.
There are two complementary scenarios of 
the secondary bifurcation, and they are described in section \ref{Multistab} and \ref{shybrid}.
In the  section \ref{sumco} we summarize our results, classify them and highlight further perspectives of this work.

\section{Generic model for a conserved order parameter field \label{genmodel}}

We introduce and investigate a nonlinear model for a real-valued conserved order parameter field $\psi(x,t)$,
which is simultaneously an extension of the Cahn-Hilliard model  \cite{Cahn:1961.1,Bray:1994.1,Desai:2009} by the generic  next higher order contributions and an extension of
the conserved Swift-Hohenberg model \cite{MatthewsPC:2000.1} by the leading order nonlinear gradient terms, 
\begin{eqnarray}
\fl \qquad
 \partial_{t}\psi = -\partial_{x}^2\Big[\varepsilon \psi+(D_4+\beta_2 \psi) \partial_{x}^2\psi -D_6\partial_{x}^4\psi
 +\beta_1\psi^2 -g\psi^3  + \beta_3 \left(\partial_{x}\psi\right)^2  \Big]\,,
\label{eq:gen_mod}
 \end{eqnarray}
 with $D_4,D_6,g>0$.   Equation~(\ref{eq:gen_mod}) is symmetric with respect to the transformation $(\psi, \beta_1, \beta_2, \beta_3) \rightarrow -(\psi, \beta_1, \beta_2, \beta_3)$. It  can be  written in the form of a conservation
 law $ \partial_{t}\psi = -\partial_x j$ with  $j=j(\psi, \partial_x \psi)$
and contains a number of models as special cases. For $D_6=\beta_1=\beta_2=\beta_3=0$ equation (\ref{eq:gen_mod}) reduces to the classic Cahn-Hilliard (CH) model in one spatial dimension \cite{Cahn:58.1,Cahn:1961.1,Bray:1994.1,Desai:2009}. Using  an $\varepsilon$-scaling  
the contributions including  $D_6,\beta_2,\beta_3$ are shown to vanish in the limit of small control 
parameter values $\varepsilon$  in  \ref{appendixCH}.

For $D_6=0$ the model corresponds to the nonlinearly extended 
Cahn-Hilliard model for active phase separation \cite{Rapp:2019.1}, which was recently derived  by a systematic perturbation calculation \cite{Bergmann:2018.2} from a dissipative mean-field model for MIPS in active colloids \cite{SpeckT:2015.1}. 
Equation (\ref{eq:gen_mod}) reduces for $D_6=\beta_1=0$ to the one-dimensional version of the so-called active model B+ with $K_1\not =0$ in reference \cite{Cates:2018.1}. 
In the case of vanishing coefficients $D_6=g=\beta_1=\beta_2=0$ the equation (\ref{eq:gen_mod}) reduces to the conserved
Kuramoto-Sivashinsky model \cite{Politi:2009.1}.
For the parameter set ${\beta_{i}=0}$ ($i=1,2,3$), ${D_6=1}$, ${D_4=-2}$ and ${\varepsilon=r -1}$  equation ~(\ref{eq:gen_mod}) 
describes the  conserved Swift-Hohenberg model for spatially periodic patterns~\cite{MatthewsPC:2000.1},
also known as  the phase-field crystal model \cite{EmmerichH:2012.1}.

The 6th order derivative $D_6 \partial_x^6 \psi$ in the linear operator in equation (\ref{eq:gen_mod}) has been added in contrast
to the models   in Refs.~\cite{Cates:2018.1,Rapp:2019.1}. Without this higher order contribution  model (\ref{eq:gen_mod})
 will become structurally unstable for larger values of the control parameter.
 Above the onset of phase separation  one has $\psi  \propto \sqrt{\varepsilon}$ and $\psi$  takes positive and negative values equally likely. 
Therefore,  the prefactor  $D_4 +\beta_2 \psi $ may become  small 
  $D_4 +\beta_2 \psi \sim   \sqrt{\varepsilon}$
for negative values of $\beta_2 \psi<0$      or even negative for $D_4>0$. Moreover,  in this parameter range close the
threshold  one has slow variations of the fields, i. e.   $\partial_x^2 \psi \sim O(\varepsilon)$,  $ \partial_x \sim O(\varepsilon^{1/4})$ and all contributions  to equation (\ref{eq:gen_mod}) are of the same order  $\propto \varepsilon^{2}$, including $\partial_x^6 \psi$.
This illustrates why equation (\ref{eq:gen_mod}) would be incomplete without the contribution  $\partial_x^6 \psi$.
In this sense equation (\ref{eq:gen_mod})  is a consistent generic model being structurally stable also for small or negative
values  $D_4 +\beta_2 \psi \sim \sqrt{\varepsilon}$. Equation (\ref{eq:gen_mod}) is a systematic extension of the conserved Swift-Hohenberg model by
leading order nonlinear gradient terms and therefore, we call it conserved Swift-Hohenberg model+ (CoSH+).

 Equation (\ref{eq:gen_mod}) has the additional interesting property that its solutions obey
a gradient dynamics  
\begin{eqnarray}
\label{grad_1}
  \partial_t \psi = \partial^2_x \left( \frac{\delta {\cal F}}{\delta \psi} \right)
\end{eqnarray}
 for the parameter combination $\beta_2=2\beta_3$   with the functional
\begin{equation}
\label{grad_2}
 {\cal F}(\psi)= \int dx \left[\frac{\psi^4}{4}  -\frac{\beta_1 \psi^3}{3}  
 -\frac{\varepsilon}{2} \psi^2  +\left( \frac{D_4}{2} + \beta_3 \psi \right) \left(\partial_x \psi\right)^2+\frac{D_6}{2}\left(\partial_x^2 \psi\right)^2    \right].
\end{equation}
In this work we focus either on the case without a gradient dynamics, 
 especially on the parameter range  $\beta_2 \propto O(1)$ with $\beta_3=0$ or small values of  $\beta_3$.
In a second case we focus on the range with a gradient dynamics and its neighborhood $\beta_2 \sim 2 \beta_3$. 
In both ranges we find two different scenarios of 
a secondary transition taking place above the onset of  phase separation. 
 
 Numerically we solve equation~(\ref{eq:gen_mod}) with periodic boundary conditions on a spatial domain $(0,L)$ (see \ref{appendixSim} for specifics). Unless stated otherwise, the following set of parameters is used throughout:  $ \beta_2=\sqrt{2}$, $D_4=0.15$, $D_6=1/6$ and $g=1$.

%
\section{Phase separation and its transition to stable  periodic patterns \label{phasepatt}}
\begin{figure}[htb]
\begin{center}
 \includegraphics[width=0.99\textwidth]{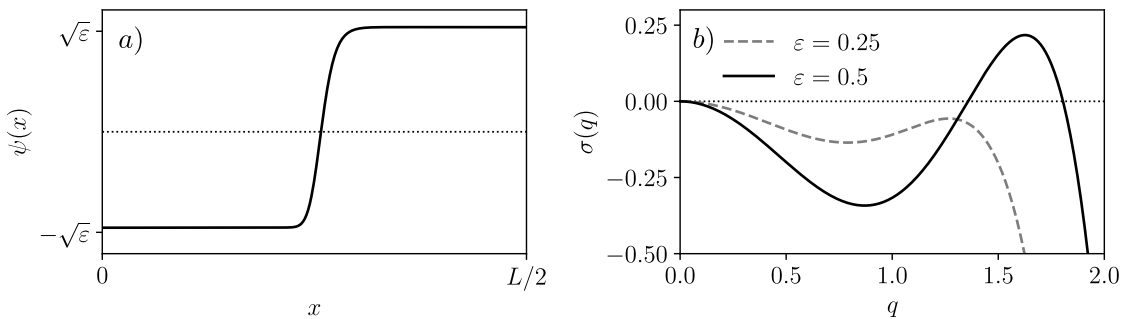}
 \end{center}
 \vspace{-6mm}
 \caption{ a) shows a one-domain solution of equation (\ref{eq:gen_mod})  for the parameters: 
 $\varepsilon=0.005$, $D_4=0.15$, $D_6=1/6$, $\beta_2=\sqrt{2}$, $g=1$, $L=500$ and $\beta_1=\beta_3=0$. 
 b) shows for two values of $\varepsilon$
  the growth rate $\sigma(q)$, cf. equation~(\ref{eq:disp}),  
 of perturbations with respect to the constant lower plateau value $\psi_\ast=-\sqrt{\varepsilon}$.
 }
 \label{fig:tanh}
\end{figure}

\begin{figure}
\begin{center}
\includegraphics[width=0.7\textwidth]{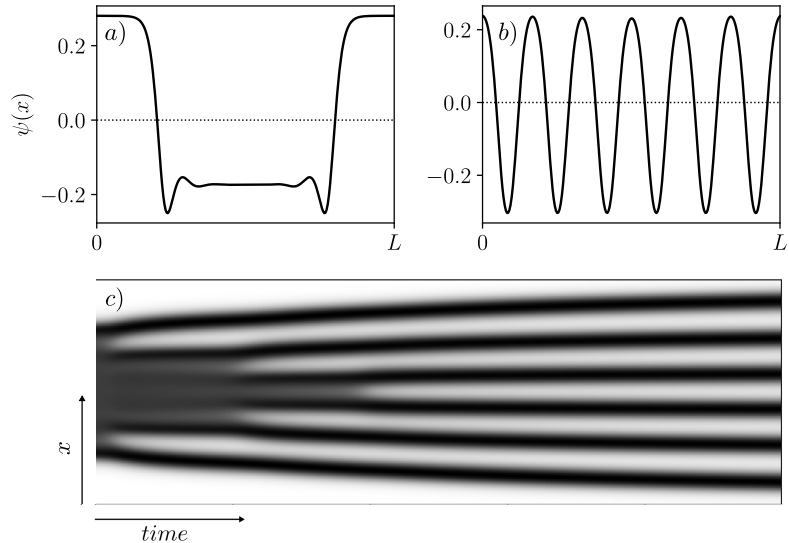}
 \end{center}
 \vspace{-1.0cm}
 \caption{Part a) shows a stationary two-domain solution of equation (\ref{eq:gen_mod}) at  $\varepsilon=0.06$ 
 and in b) a stable spatially periodic pattern  at  $\varepsilon=0.065$
The transition between the two after a jump from $\varepsilon=0.06$ to $\varepsilon=0.065$
is shown in the space-time plot in part c).
 Further parameters are $D_4=0.15$, $D_6=1/6$, $\beta_2=\sqrt{2}$, $\beta_1=\beta_3=0$, $L=100$
 and periodic boundary conditions  were used.}
 \label{fig:transition_xt}
\end{figure}

%
In the range  of small $\varepsilon$  the CoSH+ model  (\ref{eq:gen_mod}) reduces to  the classic CH-model 
as demonstrated in  \ref{appendixCH} by rescaling the CoSH+ model.  This is also true for its solutions. It 
 can be seen  in   figure~\ref{fig:tanh}a)  for  a small 
$\varepsilon=0.005$, where a  stationary solution is shown that has the same form as typical monotonic domain-wall solutions of the CH-model
 connecting the plateau values $\psi_0 = \pm \sqrt{\varepsilon\,}$ \cite{Cahn:1961.1,Bray:1994.1}.
By increasing $\varepsilon$ as in  \ref{fig:transition_xt}a) up to $\varepsilon=0.06$,  
a domain wall solution as in figure \ref{fig:tanh} is deformed  for finite $\beta_2>0$ ($\beta_2<0$) as damped oscillations develop at the lower (upper) plateau near 
the domain wall. In addition, the lower (upper) plateau widens and the amplitudes of both plateaus are shifted upwards (downwards).

To investigate whether these changes will eventually lead to an instability of domain wall solutions,
as depicted in figure 2, we use the following ansatz
\begin{eqnarray}
 \psi(x,t)=\psi_0(x)+\psi_1(x,t)\,
\end{eqnarray}
and separate the primary state, $\psi_0(x)$, from  contributions   $\psi_1(x)$
of a possible secondary state to  $\psi(x)$.
This gives the nonlinear dynamical equation for $\psi_1$:
\begin{eqnarray}
\fl \partial_t  \psi_1 = - \partial_x^2  \Big[\Big( \varepsilon  + D_4 \partial_{x}^2  - D_6 \partial_{x}^4  
- 3\psi_{0}^{2}  + 2 \beta_1 \psi_0     + \beta_2 (\partial_x^2 \psi_0 +\psi_0 \partial_x^2) 
 + 2 \beta_3 (\partial_x \psi_0)\partial_x  \Big) \psi_1\nonumber \\
  -3\psi_0\psi_1^2-\psi_1^3 +\beta_1 \psi_1^2+\beta_2\psi_1\partial_x^2 \psi_1\,
 +\beta_3 (\partial_x \psi_1)^2 \Big]\,.
\label{eq:linpsi1}
\end{eqnarray}
Further away from the domain walls and at small values of $\varepsilon$ 
the constant amplitude at the plateaus is  approximately  $\psi_\ast=\pm\sqrt{\varepsilon}$.
In a first step we neglect the spatial variation of $\psi_0(x)$ near the 
domain walls, assume at the end of the plateaus periodic boundary conditions
and the constant amplitude  $\psi=\psi_\ast$ at the whole plateau.
 In this case equation (\ref{eq:linpsi1}) simplifies with  $\beta_1=\beta_3=0$ 
to an equation with constant coefficients and its linear part can be solved 
by the ansatz $\psi_1=F \exp(\sigma t+iqx)$. The resulting growth rate has the following
$q$-dependence
\begin{equation}
 \sigma(q) = (\varepsilon-3\psi_\ast^2) q^2 -\left(D_4 +\beta_2\psi_\ast \right) q^4 -D_6q^6\,.
\label{eq:disp}
 \end{equation} 
The prefactor in front of $\propto q^4$ may
change its sign in the range $\psi_\ast \beta_2 <-D_4$
even for $D_4>0$.
In the range of a vanishing or even negative coefficient of  $q^4$ 
the stabilizing higher-order derivative $D_6\partial_x^6 \psi$ in equation (\ref{eq:gen_mod})
becomes crucial and  ensures with $D_6>0$ the damping of  short-scale perturbations in 
equation (\ref{eq:disp}).

Beyond the onset of phase separation one has $\varepsilon>0$ and therefore a negative  damping 
coefficient of  $q^2$ in equation (\ref{eq:disp}). 
In the case of a  sign change of the prefactor of $q^4$ the
growth rate $\sigma(q)$ develops besides $q=0$ a second extremum 
as indicated in figure \ref{fig:tanh}b).
The wavenumber at the finite-$q$ extremum  is
\begin{equation}
\label{eq:qmax}
 q_{max}^2=\frac{1}{3D_6}\left[-\beta_2\psi_\ast-D_4
 +\sqrt{\left(\beta_2\psi_\ast+D_4\right)^2+3D_6(\varepsilon-3\psi_\ast^2)}\right]\,,
\end{equation}
if $\psi_\ast \beta_2> (3D_6\varepsilon -D_4)$. For a sufficiently negative $D_4+\beta_2 \psi_\ast$
the growth rate becomes positive in a finite $q$-range around 
 its maximum $\sigma(q_{max})$,  as indicated by the solid line in \ref{fig:tanh}b).
In the  conserved Swift-Hohenberg model the $q$-dependence of the  perturbation growth rate  with respect to
the homogeneous basic state shows a similar pattern forming
behavior \cite{MatthewsPC:2000.1}. 

For the approximation $\psi_\ast=-\sqrt{\varepsilon}$  and $\beta_2>0$ the  critical $\varepsilon$-value, above which $\sigma(q_{max})$ becomes
positive, is given by 
\begin{equation}
\label{threshepq}
 \varepsilon_{qc} = \frac{D_4^2}{\left(2\sqrt{2D_6}-\beta_2\right)^2}\,
\end{equation}
with  $\varepsilon=\varepsilon_{qc}$  at  $q_c=q_{max}$. 
The formula  shows  that the threshold for an instability of the lower plateau value increases with
$D_4^2$. If we account for the  numerically observed upward shift in the lower plateau value,
as in  \ref{fig:transition_xt}a), by an increase of approximately  $25 \%$ compared
to  $\psi_\ast\simeq-\sqrt{\varepsilon}$, then the threshold of the instability is reduced by about $50 \%$
  according to equation  (\ref{eq:disp})  compared to $\varepsilon_{qc}$.

The spatial variation of $\psi_0(x)$ near the domain wall causes 
spatially varying coefficients in 
 the nonlinear equation (\ref{eq:linpsi1}). Such spatially varying coefficients 
 can be represented by a Fourier series
with locally varying Fourier amplitudes as demonstrated in reference \cite{Rapp:2016.1}. 
It is known, that  spatial modulations of parameters 
with a wavenumber twice that of the wavenumber of  the pattern at onset 
lead to a reduction of the pattern threshold proportional to the amplitude of the modulation 
due to a so-called $1:2$ resonance \cite{Coullet:86.2,Zimmermann:93.3}.
Here the spatially varying  coefficients in equation (\ref{eq:linpsi1}) include  Fourier modes
with a wavenumber twice that of the wavenumber $q$ of  $\psi_1$, but restricted to the range 
near the domain wall. These modulations  are  locally in
$1:2$ resonance with $\psi_1$  in conjunction with the mentioned shift of $\psi_\ast$.

In figure \ref{fig:transition_xt} a) the two-domain solution  at the control-parameter 
value $\varepsilon=0.06$ is stable. This $\varepsilon$-value is 
 far below the threshold $\varepsilon_{qc} \simeq 0.334$ 
 of an instability of a constant plateau solution $\psi_\ast =-\sqrt{\varepsilon}$ (for  parameters as in  figure \ref{fig:transition_xt}).
 However, the shift of the plateau values $\psi_\ast$ and
 the local resonance lead to a shift of the onset of a secondary pattern  such that 
 a  small $\varepsilon$-jump  to $\varepsilon=0.065$ already destabilizes 
the two-domain solution in  figure  \ref{fig:transition_xt}a)  leading to the nonlinear 
evolution as shown in figure \ref{fig:transition_xt}c).

According  to the arguments above, the domain wall solution becomes unstable far below the predicted linear stability threshold
$\varepsilon_{qc}\simeq 0.334$ for a constant plateau in the absence of spatial variations near the domain wall. 
As can be seen in figure  \ref{fig:transition_xt}c) the periodic pattern indeed nucleates  near 
the domain walls and then evolves  to a system filling extended periodic pattern
as shown in figure \ref{fig:transition_xt}b). It is quite surprising that the periodic solution 
spreads throughout the entire system instead of only over the lower plateaus.  Note that  for $\beta_2<0$ the upper plateau 
becomes unstable. The
 scenario of a periodic instability restricted 
to the lower plateau for $\beta_2>0$ (to the upper plateau for $\beta_2<0$) is described in section 
\ref{shybrid}.

\section{Multistability of spatially periodic patterns and transition scenarios \label{Multistab}}

The extended spatially periodic patterns  which displace
 active phase separation beyond the secondary instability as in figure \ref{fig:transition_xt}c), are  multistable in a finite wavenumber band, as we show in section \ref{sec:Eckh}.  In addition, we demonstrate in section \ref{sec:Transscen} that   the transitions between stable periodic patterns of different wavenumbers exhibit  the same  universal behavior as those in classical pattern forming systems 
\cite{Lowe:85.1,Zimmermann:85.1,Zimmermann:85.3,Dominguez-Lerma:86.1,Riecke:86.1,Riecke:87.1,Dominguez-Lerma:86.2,Zimmermann:88.1,Tuckerman:1990.1,CrossHo}.
Furthermore, the transition from phase separation to periodic patterns and back are shown to be hysteretic.

\subsection{The Eckhaus stability band and transitions between periodic patterns\label{sec:Eckh}} 
To address  multistability of periodic patterns of different wavenumbers we first determine
 a stationary periodic, nonlinear solution 
$\psi_s$ by using   an $N$-mode truncated Fourier ansatz 
\begin{eqnarray}
\psi_s(x) =\sum_{|\ell|\leq N} c_{\ell} e^{i \ell q x}\,,   
\label{eq:psis}
\end{eqnarray}
with the conjugate symmetry  $c_{\ell} = c^{*}_{-\ell}$ for the real field $\psi_s$ and $c_0=0$ as consequence of conservation. 
Using the ansatz (\ref{eq:psis}) and projecting equation (\ref{eq:gen_mod}) onto the modes $\exp (i n q x)$ 
for all $|n|\leq N$   yields 
coupled nonlinear equations for the Fourier amplitudes $c_n$. 
The $n$-th equation  reads
\begin{eqnarray}
\label{cn_nonlinear}
\fl \qquad 0 = L(n,0) \, c_n - \sum_{|n-m-j|\leq N} c_{n-m-j}c_m c_j
+ \sum_{|n-m| \leq N} g(n,m) \, c_{n-m}c_m\,,
\end{eqnarray}
where for  convenience we have introduced:
\begin{eqnarray}
L(n,k) &= \varepsilon - D_4 \,(k+n q)^2 - D_6 \, (k+n q)^4 \,,\nonumber \\
g(n,m) &= \beta_1 - \beta_2 \, (m q)^2 - \beta_3 \, (n-m)m q^2 \, .
\end{eqnarray}
Equations (\ref{cn_nonlinear}) are solved numerically  for a given set of system parameters and fixed values of $\varepsilon$ and  $q$.
In equation (\ref{eq:psis}) modes with even and uneven values of  $\ell$ are coupled in accordance with the broken $\pm$ symmetry of the equation
 as seen in figure \ref{fig:transition_xt}b).   The solutions of interest  are dominated mostly by the amplitude $c_{1}$. 
While still significant, higher modes decay quickly allowing us to use $N=10$ for the calculation of $\psi_s$.

Similar as  for $\psi_0(x)$ above we also  investigate the linear stability of the stationary solution  $\psi_s(x)$
with respect to a small perturbation $\psi_{p}(x,t)$. In this case one replaces in equation (\ref{eq:linpsi1})
$\psi_0$ by $\psi_s$.
With $\psi_s$  given by equation~(\ref{eq:psis}) the  part of equation being linear in $\psi_1$
 (\ref{eq:linpsi1}) can be solved with a Floquet ansatz for $\psi_1=\psi_p$,
\begin{eqnarray}
\label{eq:Floquet}
\psi_p(x,t) = e^{\sigma t} e^{i k x} \sum_{|m|\leq N} a_m e^{i m q x} \,
\end{eqnarray} 
 and real Floquet exponent $k$.
Using this ansatz  for the linear part of equation  (\ref{eq:linpsi1}) and
projecting this equation
 onto $\exp[i(k+nq)x]$,  we obtain the following linear eigenvalue problem for the coefficients $a_n$: 
\begin{eqnarray}
 \sigma a_n = &(k+nq)^2 \, L(n,k) \, a_n - 3 (k+nq)^2 \sum_{|n-m-j|\leq N} c_{n-m-j} c_{m} a_j 
\nonumber \\
& \hspace{2.5cm} + \sum_{|n-m| \leq N} h(n,m,k)  \, c_{n-m} a_m \,,
\label{eq:EWprob}
\end{eqnarray}
 where  we have introduced in addition:
\begin{eqnarray}
 h(n,m,k) = 2\beta_1 &- \beta_2 (k + nq)^2 [(k + mq)^2 + (n-m)^2q^2]  \nonumber \\
   &- 2\beta_3 (k + nq)^2 (k + mq) (n-m)q\,. 
\end{eqnarray}
By determining  the eigenvalues via equation (\ref{eq:EWprob}), we identify the parameter ranges wherein
 one finds stable periodic solutions  $\psi_s(x)$, 
i. e. wherein the eigenvalue with the largest real part $\sigma_m(k)$ is still negative:
\begin{eqnarray}
\max_{ k } \, \textrm{Re} \, \big\{ \sigma_{m}(k) \big\} < 0.
\end{eqnarray}
\begin{figure}[htb]
\begin{center}
 \includegraphics[width=0.85\columnwidth]{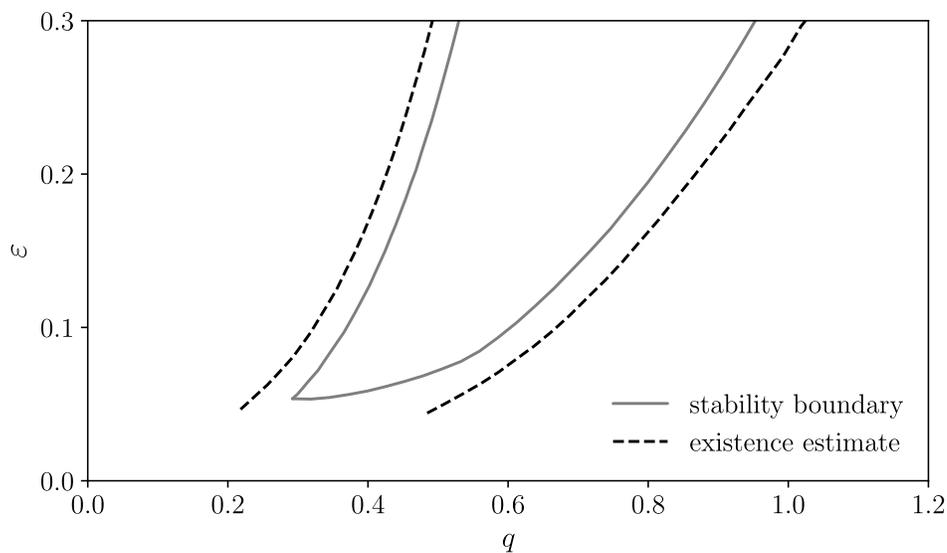}
\end{center}
\vspace{-6mm}
 \caption{
The solid line marks the boundary of the  Eckhaus stability band of  stable periodic solutions of  the generic model (\ref{eq:gen_mod}) in the $\varepsilon-q$ plane for a given parameter set as given in figure \ref{fig:tanh}.
 Unstable periodic solutions are numerically found in a wider range outside the Eckhaus band, at least in the ranges between the solid and dashed lines.
 }
 \label{fig:Eckhaus}
\end{figure}
The so-called Eckhaus stability boundary is determined by $\max_{k} \, \textrm{Re} \, \big\{ \sigma_{m}(k) \big\}=0$.
 It is displayed as the solid line in figure~\ref{fig:Eckhaus} for the parameter set given in figure \ref{fig:tanh}.
Between the Eckhaus stability boundary and the dashed lines we still  find numerically  stationary periodic solutions, but they are unstable.

Spatially periodic patterns $\psi_s$ with  wavenumbers chosen from within the Eckhaus band, i.e. the area enclosed by the solid line in
figure \ref{fig:Eckhaus}, are all linearly stable.
 This type of multistability is known to occur in a number  of other pattern forming systems beyond a primary instability
\cite{Eckhaus:65,Lowe:85.1,Zimmermann:85.1,Zimmermann:85.3,Dominguez-Lerma:86.1,Riecke:86.1,Riecke:87.1,Dominguez-Lerma:86.2,Zimmermann:88.1,Tuckerman:1990.1,CrossHo}. 
The insight that stationary, spatially periodic solutions predicted to occur beyond a secondary instability in systems with conserved order parameter fields share this generic property of multistability with a great number of other systems  \cite{Eckhaus:65,Lowe:85.1,Zimmermann:85.1,Zimmermann:85.3,Dominguez-Lerma:86.1,Riecke:86.1,Riecke:87.1,Dominguez-Lerma:86.2,Zimmermann:88.1,Tuckerman:1990.1,CrossHo}
 is central to this work.

In contrast to other systems with spatially periodic patterns the shape of the stable Eckhaus   band at its bottom in figure \ref{fig:Eckhaus} is not parabolic.
Most of the commonly known spatially periodic patterns emerge above a primary instability out of a homogeneous basic state  that is symmetric with respect to spatial reflections and translations. In these cases the Eckhaus stability band has a parabolic shape
near its minimum. Here in this work the secondary bifurcation  to a spatially periodic patterns takes place after a  primary bifurcation to phase separation and in the 
state of phase separation both symmetries are absent.

\subsection{Transition scenario to and between extended periodic patterns \label{sec:Transscen}}

\begin{figure}[htb]
\begin{center}
\includegraphics[width=0.43\columnwidth]{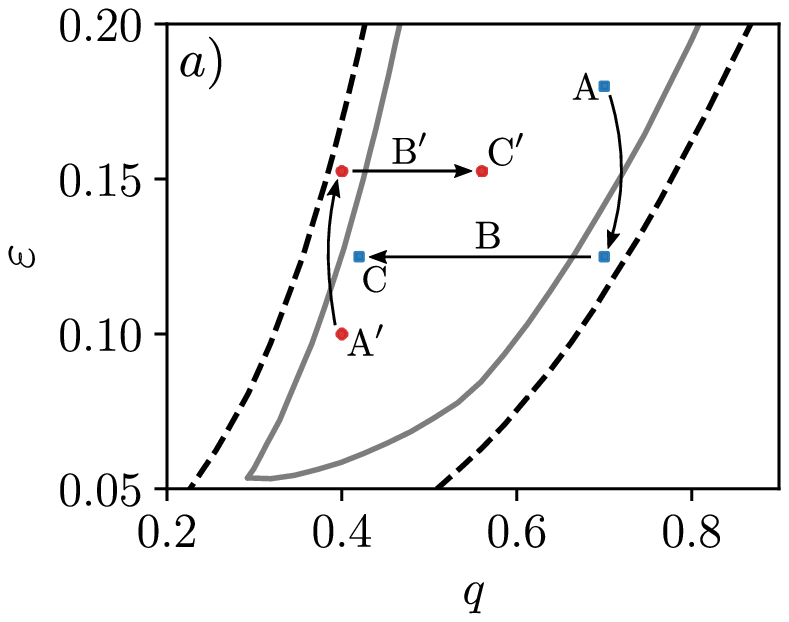}
\includegraphics[width=0.56\columnwidth]{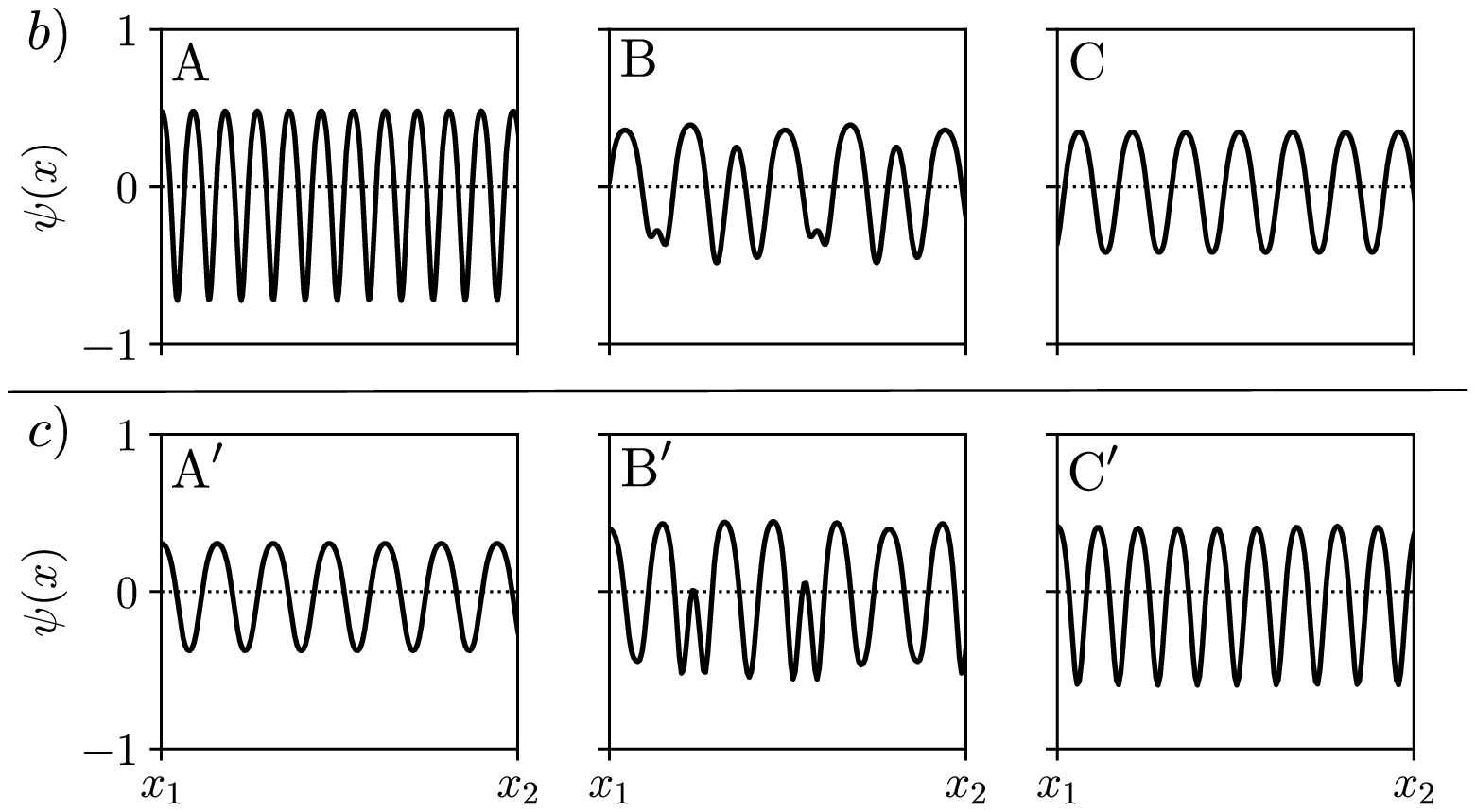}
\end{center}
 \caption{
Two typical transition scenarios between stable periodic patterns are shown. At first one starts at $A$ in the $q-\varepsilon$ plane with a stable solution $A$
 shown in  part b). We then make  a downward $\varepsilon$-jump from $A$ in the stable range in the  $\varepsilon-q$ plane  to the unstable range at $B$.
The solution evolves then  via an intermediate transient  solution as shown by $B$  in part b) to a stable periodic solution C shown in part b) 
at the parameters marked by $C$ in part a). In the second scenario we start with a stable solution $A'$ shown in part c)  at the parameters   
     $(q,\varepsilon)$  marked by$A'$ in a). 
After an upward $\varepsilon$ jump from the stable $A'$ to the unstable point $B'$ in part a)  the stable solution $A'$ in c) 
evolves via transients as $B'$ in part c)  to a stable solution $C'$ in c)  at the parameters $C'$ in a).
 }
 \label{fig:Eckhaus_dyn}
\end{figure}

The similarities to other nonequilibrium stripe pattern are highlighted further 
in the way the solutions of the CoSH+ model in equation (\ref{eq:gen_mod}) respond to changes in the control parameter or the wavenumber
 beyond the secondary instability.
 This is demonstrated in figure~\ref{fig:Eckhaus_dyn} for a generic property of periodic patterns.

Figure \ref{fig:Eckhaus_dyn}a) depicts two examples of transition  scenarios: In the first one, $ A \to C$, a
 linearly stable periodic solution inside the Eckhaus band is quenched to the unstable range between the band and existence estimate.
The system reacts to this change by reducing the  wavenumber of the pattern via a series of transient states $B$ to arrive at $C$ back inside the Eckhaus band.
 Figure  \ref{fig:Eckhaus_dyn}b)   shows the corresponding initial stable state $A$, an example of a transient $B$ and the final state
$C$. In the second scenario  $A'\to C'$, we jump upwards into the unstable range and the
system reacts by increasing the pattern wavenumber via transients $B'$  to arrive at $C'$  within the band. Figure   \ref{fig:Eckhaus_dyn} c) 
again shows the corresponding initial stable, the final stable periodic solution and an example of a transient $B'$. Note that in both scenarios the shift
 in $\varepsilon$ is accompanied by a shift in the amplitude, which increases and decreases with
the control parameter.

 These transition scenarios
of periodic states after parameter changes as shown in figure \ref{fig:Eckhaus_dyn} 
are also observed for spatially periodic patterns in models with unconserved order parameters \cite{Zimmermann:85.1}, in  
electrohydrodynamic convection
in nematic liquid crystals 
\cite{Lowe:85.1} and in experiments   with periodic patterns in Taylor-vortex flows \cite{Dominguez-Lerma:86.1}.
Note  that in these examples the transition scenarios  have been
observed for patterns that emerge out of a homogeneous basic state.
In contrast, periodic patterns  in our example  emerge  from  a phase-separated state  with
broken translational  and  the reflection symmetry.  Nevertheless, the
transition scenarios for periodic patterns in
a conserved model as shown in figure \ref{fig:Eckhaus_dyn} 
  are generic  as  in all the previous examples.

\begin{figure}
\begin{center}
 \includegraphics[width=0.85\columnwidth]{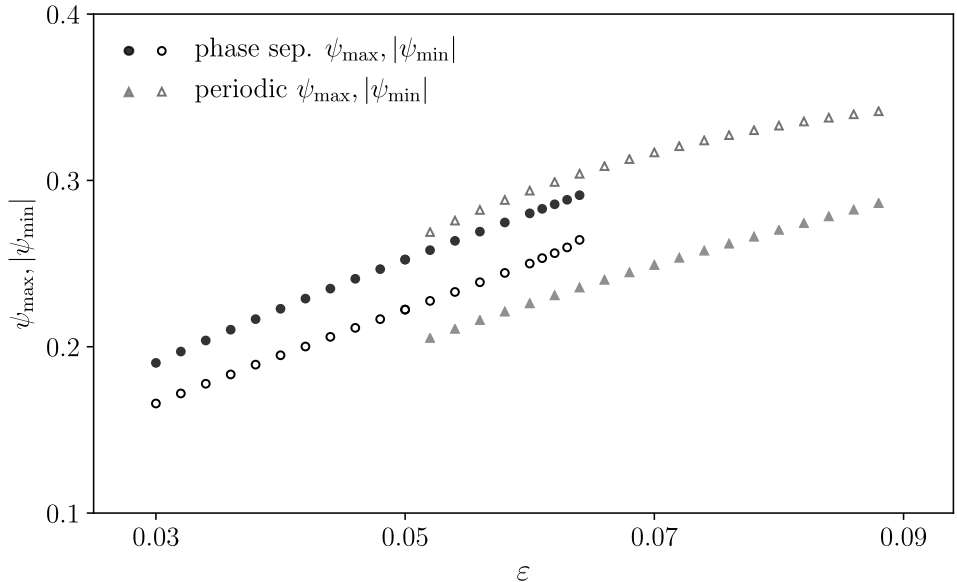}
\end{center}
 \vspace{-6mm}
 \caption{The maximum value $\psi_{max}$ and the magnitude of the minimal value $|\psi_{min}|$ of steady state solutions
 in simulations of equation~(\ref{eq:gen_mod}) are given as a function of $\varepsilon$: Circles correspond
 to the amplitude $\psi_{max}$ (filled circles) and $|\psi_{min}|$ (open circles)  during phase separation and the triangles mark 
 the amplitudes  of periodic solutions.
 If one starts with a state of phase separation the transition
 to spatially periodic solutions takes place by increasing $\varepsilon$ at around $\varepsilon\approx 0.065$
for the parameters given in figure \ref{fig:tanh}. Starting with periodic solution, this will be stable  below 
$\varepsilon\approx 0.065$, i. e. the secondary transition between phase separation and periodic patterns is hysteretic.
}
 \label{fig:transition_eps}
\end{figure}

Due  to the quadratic nonlinearities including $\beta_2$ and $\beta_3$ in equation (\ref{eq:gen_mod}),  the 
maximal amplitudes $\psi_{max}$ and $|\psi_{min}|$ differ in both   the phase separated state and the periodic state, cf. figure \ref{fig:transition_xt}a),b).
Plotting them as functions of increasing $\varepsilon$ one finds at the transition from phase separation (circles) to the periodic state (triangles) a jump
in both amplitudes as shown in figure \ref{fig:transition_eps}.
 Repeating the same by starting with a periodic pattern and decreasing  $\varepsilon$ one finds the hysteresis of the transition 
as shown in figure \ref{fig:transition_eps}.

By increasing $\varepsilon$, the transition scenario from phase separation to periodic states 
is similar as shown in figure~\ref{fig:transition_xt}c). The wavenumber of the final state
is always within the Eckhaus-stable band in figure \ref{fig:Eckhaus}, whereby the wavenumber of the final periodic state
may depend on the spatial structure of phase separation, i. e. on the distribution of the domain walls  just before the transition takes place.

Starting with a periodic pattern, a sufficient reduction of $\varepsilon$ leads latest at the lower boundary of
the Eckhaus stability band 
to a transition to phase separation. Since the spatial structure of the state of phase separation is
multifaceted in long systems, the transition dynamics from periodic pattern to phase separation is in general less universal and rather complex.

The hysteretic transition scenario shown in figure  \ref{fig:transition_eps} for $\beta_1=\beta_3=0$  will not change very much for finite
 but not to large
parameter values  $\beta_3$ and $\beta_1$. However, in the range $\beta_2 \sim 2\beta_3$ we find a different 
transition scenario as described in the next section \ref{shybrid}.

 \section{Stable secondary hybrid states \label{shybrid}}

For the parameter combination $\beta_2=2\beta_3$  model  (\ref{eq:gen_mod}) has the surprising property 
of a gradient dynamics as described by equations (\ref{grad_1}) and (\ref{grad_2}). 
The stability analysis of a constant plateau solution in section~\ref{phasepatt} is independent of $\beta_3$ and so the threshold $\varepsilon_{qc}$ 
remains unaltered.  The dynamics and stationary solutions above  $\varepsilon_{qc}$,    however, differ greatly from those described in sections~\ref{phasepatt} and \ref{Multistab} for $\beta_3=0$.  Note that for $\beta_2<0$ the upper instead of  of the lower plateau becomes unstable, but the scenario is otherwise unchanged.

\begin{figure}[htb]
 \begin{center}
 \includegraphics[width=0.85\textwidth]{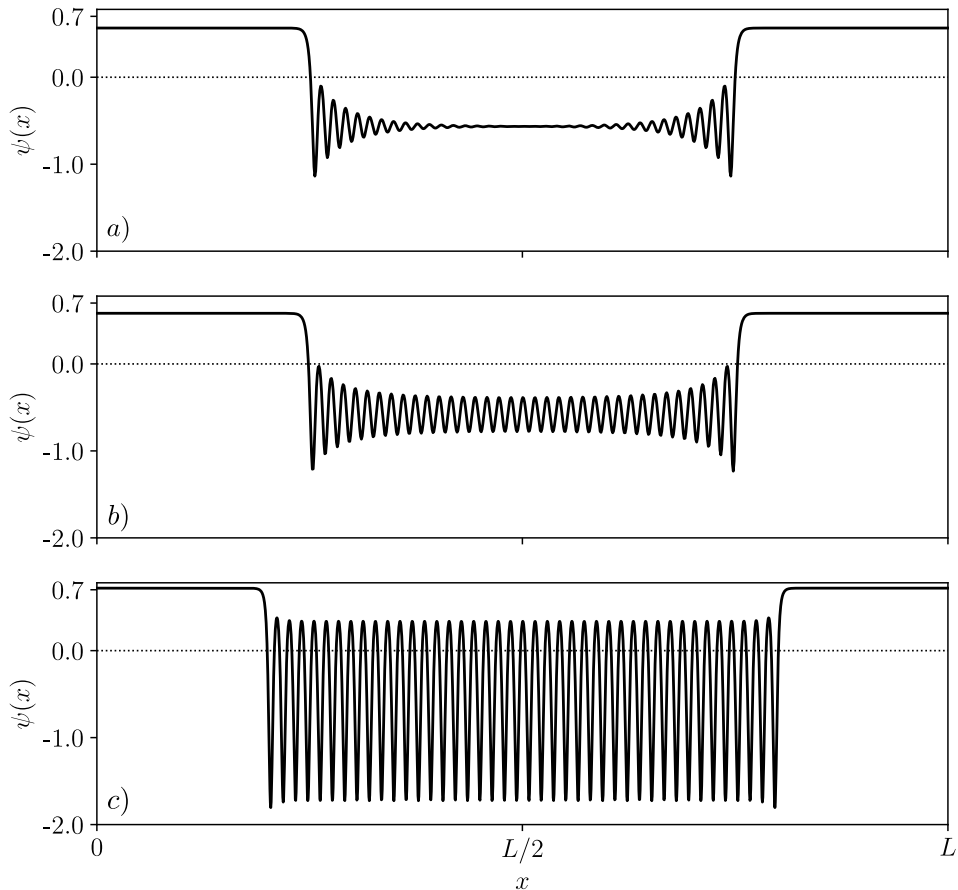}
 \end{center}
 \vspace{-6mm}
 \caption{
 Stationary solutions of equation (\ref{eq:gen_mod}) are shown
  for three different values of the control parameter  $\varepsilon=0.32$ in a), $\varepsilon=0.34$ in b) and  
 $\varepsilon=0.5$ in c) for the parameter combination  $\beta_2=2\beta_3=\sqrt{2}$ and $\beta_1=0$, 
 where the solutions of equation (\ref{eq:gen_mod})
follow according to equations (\ref{grad_1}) and (\ref{grad_2}) a gradient dynamics. The length
is $L=320$  and further parameters as in figure \ref{fig:transition_xt}.
 }
 \label{fig:hybrid}
\end{figure}

We find in a range around  $\beta_2 \sim 2\beta_3$ a further 
  novel transition from the state of phase separation, with many plateau areas of different length, 
  to a hybrid state, where constant plateaus alternate with spatially periodic patterns,  as shown for example in a shorter system with only 
two domain walls in  figure \ref{fig:hybrid}.  
Similar  areas  are occupied by the plateaus  and the periodic states  according to the conservation of the field $\psi(x,t)$, but
by increasing the distance $\varepsilon -\varepsilon_{qc}$, the region occupied by a periodic state increases slightly
as indicated in figure \ref{fig:hybrid} from part b)  to c).

In the range  $\beta_2 \simeq 2\beta_3$ the secondary periodic pattern  occurs again at first locally  near the domain walls via  wavy
 overshoots caused by the higher order spatial derivative.
Therefore,  localized finite amplitude patterns  occur already below the threshold $\varepsilon_{qc}$, 
as indicated in figure \ref{fig:hybrid}a)
at $\varepsilon=0.32< \varepsilon_{qc}=0.334$ (for the parameters used), but do not invade the lower plateau range.
Patterns localized near the domain walls 
in the range $\varepsilon  \lesssim \varepsilon_{qc}$, as in figure \ref{fig:hybrid}a), 
decay exponentially with the distance from the domain walls and the decay length 
increases with decreasing values of $D_6$. 

A numerical solution of equation (\ref{eq:gen_mod}) shows that the   amplitude of the periodic pattern in the middle 
between the two domain walls  increases or decreases
 continuously  $ \propto \sqrt{\varepsilon-\varepsilon_{qc}}$
by increasing or decreasing $\varepsilon$. Therefore, the transition  from the state of phase separation to the hybrid state is continuous and  
 not hysteretic in spite of its local onset 
near the  domain walls.

 We have confirmed this  numerical observation by a two-mode approximation $\psi_{hyb}=\psi_\ast + A \cos(q_{max}x)$. By choosing 
this ansatz we   assume a
long lower plateau at $\psi_\ast$ and periodic boundary conditions at its ends, i. e. we neglect the effects of the domain walls at the ends
 of a long plateau.
In this case we obtain in agreement with our numerical observation that the amplitude $A$ is real with $A^2>0$ only in the range
of $\varepsilon>\varepsilon_{qc}$.
 One finds $\psi_{hyb}$ with  $A\not =0$ only in the range $\varepsilon > \varepsilon_{qc}$. The  constant plateaus with $A=0$ 
corresponds to a local maximum, i. e.  the functional has for finite $A$ a lower value than for $A=0$.

\section{Summary and discussions\label{sumco}}
  
We discovered two new secondary bifurcation scenarios  in systems described by conserved order parameter fields. Both scenarios occur by increasing the control parameter further after the primary bifurcation into active phase separation.
In  the first case, phase separation is stopped and completely replaced by an extended, system-filling, spatially periodic pattern. 
It is stable for different values of the wavenumber  that belong to the  so-called Eckhaus stability band,  
which we have  determined for our model   as shown in figure \ref{fig:Eckhaus}.
The  multistability of patterns with different wavenumbers  is universal   
  \cite{Lowe:85.1,Zimmermann:85.1,Zimmermann:85.3,Dominguez-Lerma:86.1,Riecke:86.1,Riecke:87.1,Dominguez-Lerma:86.2,Zimmermann:88.1,Tuckerman:1990.1,CrossHo}.
It has been found experimentally and theoretically in a considerable number of nonequilibrium pattern forming systems with  unconserved order parameter fields
\cite{Lowe:85.1,Zimmermann:85.1,Zimmermann:85.3,Dominguez-Lerma:86.1,Riecke:86.1,Riecke:87.1,Dominguez-Lerma:86.2,Zimmermann:88.1,Tuckerman:1990.1,CrossHo}. 
As we have shown here, this multistability also applies 
for periodic patterns beyond a secondary bifurcation in  systems with a conserved order parameter field.    

The second bifurcation scenario  describes a transition   from   active phase separation to a novel and stable
 nonlinear hybrid state. 
These hybrid solutions are spatial alternations between sections of spatially constant and spatially periodic states.
Surprisingly, this spatially alternating coexistence of different spatial structures is stable.
The subranges occupied by the  homogeneous and periodic states
may vary in space, but due  to the conservation law both occupy similar areas immediately above the transition to  the hybrid state.
With increasing distance of the control parameter from the secondary threshold, the regions occupied by the periodic states   increases slightly.

 The basis of our novel results is the CoSH+ model in equation (\ref{eq:gen_mod}), which we introduced with this work.
 This model is a  further step within a   recent systematic and dynamic  development for  nonequilibrium  systems described by conserved order parameter fields that we mention shortly. It has been found recently, that the weakly nonlinear behavior of 
several demixing phenomena  in systems far from equilibrium  are also described  at leading order by the 
 classic Cahn-Hilliard equation \cite{Bergmann:2018.2,Rapp:2019.1,Bergmann:2019.1,Rapp:2019.2} .
 This was demonstrated exemplarily for collective demixing phenomena in chemotactic systems \cite{Bergmann:2018.2,Rapp:2019.2}, for models of cell polarization \cite{Bergmann:2018.2,Bergmann:2019.1} and motility induced phase separation (MIPS) in active Brownian particle 
 systems \cite{Rapp:2019.1}.  Previously, for modeling MIPS further away from its onset  an extension of the CH model to the next higher nonlinear  order,
i. e.  $\partial_x^2(\psi \partial_x^2 \psi)$ and $\partial_x^2(\partial_x \psi)^2$, was suggested with the introduction of 
active model B+ \cite{Cates:2014.1,Cates:2018.1}.
 One of the two nonlinear extensions contributes to the prefactor
 $D_4 +\beta_2\psi$  of $\partial_x^4 \psi$ in the active model B+   and  in the CoSH+ model (\ref{eq:gen_mod}).
 This prefactor  may change its sign   at larger amplitudes of the order parameter field.
In this range the contribution  $D_6 \partial_x^6 \psi$ is according
to our perturbation theory of the same order  as the nonlinear extensions.   It is indispensable in this range to avoid 
dangerous large wavenumber instabilities after sign change     of $D_4 +\beta_2\psi$.  
Our model  in (\ref{eq:gen_mod}) includes the sixth order derivative, similar  as the conserved 
Swift-Hohenberg  (CSH) model in reference \cite{MatthewsPC:2000.1}, and it is  in this sense an extension 
of the conserved SH model by generic leading order nonlinear gradient terms. Accordingly,   we call it in section \ref{genmodel} the conserved Swift-Hohenberg model+ (CoSH+ model).
  
The spatially varying sign change of the prefactor $D_4 +\beta_2\psi$    in equation (\ref{eq:gen_mod})
is the origin of the two novel bifurcation scenarios. This sign change 
 is qualitatively similar to the Lifshitz point in pattern formation, found near the so-called oblique-roll instability in electroconvection in nematic liquid
crystals, where  the coefficient of a higher order spatial
derivative undergoes a similar change in sign, but spatially homogeneously \cite{Zimmermann:85.2,Ribotta:86.1,Pesch:86.1,Zimmermann:88.3}.

Secondary bifurcations of, for example,  stripe patterns  are  known in the field of classical pattern formation  with unconserved order parameter fields
\cite{Busse:78.1,Coullet:90.1,Coullet:91.1,BoPeAh:2000.1,Misbah:94.3,Karma:1996.1,Faivre:97.1,CrossHo}. 
However,  to   the best of our knowledge there is so far
no further example of a secondary bifurcation that evolves   
in a conserved nonequilibrium system  out of  active phase separation into  stable nonlinear periodic pattern.

The existence of  stability bands of periodic patterns, which we found for our first secondary bifurcation scenario,  is  a universal and robust phenomenon in nonlinear physics.     To add or to  remove a periodic unit (wavelength) of a pattern
  requires  strong local deformations of the pattern, which cannot be driven by small inherent  fluctuations. Therefore, in the absence of strong external deformations,
periodic patterns remain stable at a wavenumber within the Eckhaus band. It is an important result of our work that this robust behavior of periodic patterns applies also  to patterns beyond the secondary bifurcation in conserved systems. 
This robust stability of periodic patterns in conserved systems  is also
highly relevant for modeling  cell division, where it possibly helps  to find the center of a cell with great certainty, 
before it divides into two equally sized daughter cells carrying the same genetic information \cite{Bergmann:2018.1,Sourjik:2017.1}.

 In some former studies active phase separation was changed  into a bifurcation to spatially periodic patterns  by the introduction of additional  kinetic effects that violate conservation laws and therefore change the bifurcation type \cite{Fromherz:95.1,Ziebert:2004.1,Cates:2010.1}.
 For such a kinetically changed   primary bifurcation from phase separation to periodic patterns,  the notion arresting phase separation was coined \cite{Cates:2010.1}. Note that this is a crucial difference between these examples and our work. The transition to periodic patterns occurs in our work via a secondary bifurcation  under the retention of the conservation conditions.

An extension  of the model in equation (\ref{eq:gen_mod}) to two spatial dimensions is a further promising perspective.   It is 
 very likely that our results obtained for one spatial dimension  lead in two dimensions to further interesting  secondary bifurcation scenarios,
for example,  due to the broken up-down symmetry one expects at even lower values of the control parameter 
a secondary bifurcation to hexagonal patterns.

For pattern forming systems with unconserved order parameters  a number of interesting finite
size,  inhomogeneity,  noise  and  boundary effects  on periodic patterns are known
\cite{Greenside:84.1,BoPeAh:2000.1,Coullet:86.2,Zimmermann:93.3,Freund:2011.1,Kaoui:2015.1,Rapp:2016.1,Bergmann:2018.1,Ruppert:2020.1}. In contrast, such effects on 
periodic patterns in conserved systems, including those beyond secondary bifurcations, are unexplored so far.  

Candidates of specific systems where the  described  phenomena  are expected as well, are, for example,  generalized chemotactic systems
(see e.\,g. \cite{Rapp:2019.2}),  Brownian particle systems showing motility induced phase separation in the presence
of quorum  sensing \cite{SpeckT:2020.2} 
 and cell biology, where periodic patterns play a key role in finding the cell 
center so that cells divide into equal parts with the same amount of genetic information \cite{Sourjik:2017.1}.

 \ack
Support by the Elite Study Program Biological Physics and discussions with
Andre F\"ortsch, Samuel Grimm, Markus Hilt, Mirko Ruppert and Winfried Schmidt are gratefully acknowledged.

\appendix

\section{Significance of higher order corrections}
\label{appendixCH}
In this appendix,  we show as in reference \cite{Rapp:2019.1} that contributions to our model (\ref{eq:gen_mod}), which go
  beyond the classic Cahn-Hilliard model, vanish  in the limit of small values of 
the distance $\varepsilon$ from the onset of phase separation. 
For this  we rescale time, space and amplitude in equation~(\ref{eq:gen_mod}) such that the part of the  classic CH in 
equation  (\ref{eq:gen_mod}) becomes dimensionless: 
 $t'= \varepsilon^2 t$, $x'= \sqrt{\varepsilon} x$  and 
$\psi'=\psi/\sqrt{\varepsilon} $. In addition we choose $\beta_1=\varepsilon \beta_1'$, because the quadratic term
at leading order can be removed by adding a constant. 
This allows us to rewrite equation~(\ref{eq:gen_mod}) with a dimensionless classic CH part  in the following form:
\begin{eqnarray}
 \partial_{t'}\psi' &=& -\partial_{x'}^2\left[ \psi'+\partial_{x'}^2\psi'-\psi'^3 \right]\nonumber\\
 &&+  \sqrt{\varepsilon} \partial_{x'}^2  \left[\beta_2 \psi' \partial_{x'}^2\psi'  + \beta_3 \left(\partial_{x'}\psi'\right)^2+\beta_1' \psi'^2\right] 
+ \varepsilon \partial_{x'}^6 \psi'\,.
\label{eq:amp_rescaled}
 \end{eqnarray}
Therefore,  all contributions which do not belong to the leading order classic Cahn-Hilliard model vanish in the limit 
of  small $\varepsilon$. This behavior explains, why in figure \ref{fig:tanh}a) the solution of model (\ref{eq:gen_mod}) has
in  the limit of small $\varepsilon$ the same form as domain wall solutions of  the classic CH model. 

\section{Simulation method} \label{appendixSim}
We perform spatio-temporal simulations of equation~(\ref{eq:gen_mod}) on a spatial domain $(0,L)$ with periodic boundary conditions. We focus on stationary solutions and use a pseudo-spectral method with semi-implicit Euler time-stepping, treating the linear differential operator implicitly and the nonlinear contributions explicitly. The interval is discretized uniformly into an $N$-point grid with resolution $\Delta x = L/N$, where $N$ (i.e. the number of Fourier modes) is adapted according to the number of periods per system length. For Figures~\ref{fig:transition_xt},\ref{fig:Eckhaus_dyn} and \ref{fig:transition_eps} we used $L=100,N=256$ with time-step size $\Delta t = 1 \times 10^{-3}$. To resolve the more rapid spatial variations in the functional case displayed in Figure~\ref{fig:hybrid} we used $N=1024$ with $\Delta t = 1 \times 10^{-4}$. 
 
\vspace{5mm}

\providecommand{\newblock}{}


 
\end{document}